\documentclass[12pt]{article} 
\usepackage{bbold}
\usepackage{cite}
\usepackage{graphicx}
\def \bea{\begin{eqnarray}} 
\def \beq{\begin{equation}}

\def \eea{\end{eqnarray}} 
\def \eeq{\end{equation}}

\def \s{\sqrt{2}} 
\def \st{\sqrt{3}} 
\def \sx{\sqrt{6}} 
\def\lsim{\mathrel{\rlap{\lower3pt\hbox{$\sim$}}\raise2pt\hbox{$<$}}}
\def\gsim{\mathrel{\rlap{\lower3pt\hbox{$\sim$}}\raise2pt\hbox{$>$}}}

\begin{document} 
\begin{flushright}
EFI 15-39 \\
TECHNION-PH-2015-14 \\
arXiv:1512.06700 \\
December 2015 \\
\end{flushright} 
\centerline{\bf S-WAVE NONLEPTONIC HYPERON DECAYS AND $\Xi^-_b \to \pi^-
\Lambda_b$}
\medskip
\centerline{Michael Gronau}
\centerline{\it Physics Department, Technion, Haifa 32000, Israel}
\medskip 
\centerline{Jonathan L. Rosner} 
\centerline{\it Enrico Fermi Institute and Department of Physics,
  University of Chicago} 
\centerline{\it Chicago, IL 60637, U.S.A.} 
\bigskip

\begin{quote}
The decay $\Xi^-_b \to \pi^- \Lambda_b$ has recently been observed by the
LHCb Collaboration at CERN.  In contrast to most weak decays of $b$-flavored
baryons, this process involves the decay of the strange quark in $\Xi_b$,
and thus has features in common with nonleptonic weak decays of hyperons.
Thanks to the expected pure S-wave nature of the decay in question
in the heavy $b$ quark limit, we find
that its amplitude may be related to those for S-wave nonleptonic decays
of $\Lambda$, $\Sigma$, and $\Xi$ in a picture inspired by duality.  The
calculated branching fraction ${\cal B}(\Xi^-_b \to \pi^- \Lambda_b) = (6.3
\pm 4.2) \times 10^{-3}$ is consistent with the range allowed in the
LHCb analysis.  The error is dominated by an assumed 30\% uncertainty in
the amplitude due to possible U(3) violation.  A more optimistic view
based on sum rules involving nonleptonic hyperon decay S-wave amplitudes
reduces the error on the branching fraction to $2.0 \times 10^{-3}$.
\end{quote}

\leftline{\qquad PACS codes: 14.20.Mr, 14.20.Jn, 12.40.Nn, 13.30.Eg}
\bigskip

\section{INTRODUCTION} \label{sec:intro}

Most decays of $b$-flavored baryons observed up to now occur with the $b$
quark decaying to $c$ or $u$ and a virtual $W^-$, or via a $b \to s$ or $b \to
d$ penguin amplitude. However, it has long been noted that strange $b$-flavored
baryons are heavy enough that their $s$ quarks can decay instead via the
subprocess $s \to \pi^- u$ or $su \to ud$, with the $b$ quark acting as a
spectator \cite{Cheng:1992ff,Sinha:1999tc,Voloshin:2000et,Li:2014ada,
Faller:2015oma,Cheng:2015ckx}.
Such processes enable the decays $\Xi^{(-,0)}_b \to \pi^{(-,0)} \Lambda_b$.
The strange quark in a charmed-strange baryon can also decay to a nonstrange
one.  Here the additional subprocess $c s \to d c$ can contribute.  Thus the
decays $\Xi^{(+,0)}_c \to \pi^{(+,0)} \Lambda_c$ are permitted.

The LHCb Collaboration at CERN has now observed the decay $\Xi^- \to \pi^-
\Lambda_b$ \cite{Aaij:2015yoy}.  The quantity measured is the product of
the ratio of fragmentation functions $f_{\Xi^-_b}/f_{\Lambda_b}$ and the
branching fraction ${\cal B}(\Xi^-_b \to \pi^- \Lambda_b)$:
\beq
\frac{f_{\Xi^-_b}}{f_{\Lambda_b}}{\cal B}(\Xi^-_b \to \pi^- \Lambda_b)
= (5.7 \pm 1.8 ^{+0.8}_{-0.9}) \times 10^{-4}~.
\eeq
Assuming a range $0.1 \le f_{\Xi^-_b}/f_{\Lambda_b} \le 0.3$ based on measured 
production rates of other strange particles relative to their non-strange counterparts, the
branching ratio is then expected to lie between $(0.57 \pm 0.21)\%$ and
$(0.19 \pm 0.07)\%$~\cite{Aaij:2015yoy}.

This decay is expected to proceed purely via S-wave. As a result, we find in a
framework inspired by duality that its amplitude may be related to those for
S-wave nonleptonic decays of $\Lambda$, $\Sigma$, and $\Xi$.  The 
calculated branching fraction ${\cal B}(\Xi^-_b \to \pi^- \Lambda_b) = (6.3 \pm 4.2)
\times 10^{-3}$ is consistent with the range allowed in the LHCb analysis.

Earlier studies of $\Xi^-_b \to \pi^- \Lambda_b$ (and other 
heavy-flavor-conserving heavy baryon
decays) \cite{Cheng:1992ff,Sinha:1999tc,Voloshin:2000et,Li:2014ada,Faller:%
2015oma,Cheng:2015ckx} have applied current algebra and a soft pion limit,
thereby relating this amplitude to a matrix element of a strangeness-changing
four-fermion operator between the initial and final heavy baryon states.
Calculations of this matrix element are model-dependent and involve large
theoretical uncertainties.

The S-wave nature of the decay $\Xi^-_b \to \pi^- \Lambda_b$ is discussed
in Sec. \ref{sec:swa}.  A duality-inspired description of the S-wave decays
of the strange hyperons $\Lambda$, $\Sigma$, and $\Xi$ is recalled and updated
in Sec.\ \ref{sec:nlh}.  As the bottom quark acts as a spectator in $\Xi^-_b
\to \pi^- \Lambda_b$, we relate the amplitude for decay of its strange quark
to S-wave amplitudes of the strange hyperons in Sec.\ \ref{sec:rel}.  The
parameters extracted from an updated analysis of hyperon S-wave amplitudes are
then applied to calculate the rate for $\Xi^-_b \to \pi^- \Lambda_b$ in Sec.\
\ref{sec:pred}.  A more optimistic view of possible uncertainties is presented
in Sec.\ \ref{sec:opt}.  Section \ref{sec:con} concludes.

\section{S-WAVE NATURE OF THE DECAY $\Xi^-_b \to \pi^- \Lambda_b$}
\label{sec:swa}

In the $\Lambda_b = b[ud]$, the light quarks $u$ and $d$ are in an S-wave
state with $I = S = 0$ (denoted by the square brackets).  To the extent that
$|m_s - m_d|$ can be neglected in comparison with $m_b$ \cite{Maltman:1980er,%
Rosner:1981yh,Karliner:2008sv}, the light quarks $s$ and $d$ in
$\Xi^-_b$ also are in an S-wave state anstisymmetric in flavor with $S = 0$.
In the decay $\Xi^-_b \to \pi^- \Lambda_b$, the $b$ quark acts as a spectator.
The transition among light quarks is thus one with $J^P = 0^+ \to \pi 0^+$,
and hence is purely a parity-violating S wave.  We shall see that it thus may
be related to parity-violating S-wave amplitudes in the nonleptonic decays of
$\Lambda$, $\Sigma$, and $\Xi$.

\section{S-WAVE NONLEPTONIC HYPERON DECAYS} \label{sec:nlh}

Many attempts have been made to systematize nonleptonic weak decays of the
hyperons belonging to the lowest SU(3) octet.  For discussions of both
parity-violating S-wave and parity-conserving P-wave decays, see, e.g.,
Refs.\ \cite{Gronau:1972pj,Gronau:1972pa,Wu:1985yb} and numerous references
therein.  The P-wave decays are sensitive to delicate cancellations among
pole contributions.  In contrast, there is a compact parametrization of the
S-wave decays \cite{Nussinov:1969hp} based on Dolen-Horn-Schmid duality
\cite{Dolen:1967zz,Dolen:1967jr} which does a 
very good job in describing the parity-violating amplitudes. 
We will adopt this approach, which has been shown to be equivalent to earlier 
studies of these S-wave amplitudes using current algebra and partial 
conservation of the axial current (PCAC) 
and assuming octet dominance~\cite{Sugawara:1965zza,Suzuki:1965zz}.

We recall the notation of Refs.\ \cite{Wu:1985yb,Roos:1982sd}.  With the
effective Lagrangian for the decay given by
\beq
{\cal L}_{\rm eff} = G_F m_\pi^2[\bar \psi_2(A + B \gamma_5)\psi_1] \phi_\pi~,
\eeq
the partial width for $B_1 \to \pi B_2$ is
\beq \label{eqn:rate}
\Gamma(B_1 \to \pi B_2) = \frac{(G_F m_\pi^2)^2}{8 \pi m_1^2} q
 [(m_1+m_2)^2-m_\pi^2]|A|^2 + [(m_1-m_2)^2-m_\pi^2]|B|^2~.
\eeq
Here $G_F = 1.16638 \times 10^{-5}$ GeV$^{-2}$ is the Fermi decay constant,
$q$ is the magnitude of the final three-momentum of either particle in the
$B_1$ rest frame, and $A$ and $B$ are S-wave and P-wave decay amplitudes,
respectively. 

The nonleptonic hyperon decays are observed to follow an approximate
$|\Delta I| = 1/2$ rule, so a decay $B_1 \to \pi B_2$ may be regarded as a
scattering process $\sigma B_1 \to \pi B_2$, where the spurion $\sigma$
transforms as a $K^0$ and has $J^P = 0^-$ for parity-violating S-wave
amplitudes and $0^+$ for parity-conserving P-wave amplitudes.  In Ref.\
\cite{Nussinov:1969hp}
it was noted that one could regard the S-wave amplitudes as if the
spurion and pion coupled to a $t$-channel $K^*$ or $K^{**}$ Regge trajectory,
with relatively small [{\cal O}(15\%)] contributions from octet pole terms.
Such a Regge-dominance picture would not hold for the pole-dominated P-wave
amplitudes.  With the assumption that all S-wave decays can be regarded as
if a $K^0 \pi$ pair were exchanging a Regge trajectory with the $\bar B_1 B_2$
pair, one can calculate all S-wave $B_1 \to \pi B_2$ amplitudes in terms of
an overall scale $x$ and a parameter $F$ expressing the ratio of antisymmetric
to symmetric three-octet coupling at the baryon vertex.  In our convention,
$F + D = 1$.  A typical graph illustrating the coupling in $\Sigma^- \to \pi^-
n$ is shown in Fig.\ 1(a).

An exchanged trajectory $E$ and the baryons $B_1$ and $B_2$ may be represented
by $3 \times 3$ matrices.  For $\pi^-$ emission, $E$ represents the product
of $\bar \sigma$ (transforming as a $\bar K^0$) and the $\pi^-$ matrix, or
\beq
E = \left[ \begin{array}{c c c} 0 & 0 & 0 \cr 0 & 0 & 0 \cr 1 & 0 & 0
  \end{array} \right]
\eeq
For octet baryons,
\beq
B = \left[ \begin{array}{c c c} \frac{\Sigma^0}{\s} + \frac{\Lambda}{\sx} &
\Sigma^+ & p \cr \Sigma^- & -\frac{\Sigma^0}{\s} + \frac{\Lambda}{\sx} &
n \cr \Xi^- & \Xi^0 & -\frac{2 \Lambda}{\sx} \end{array} \right]~,~~ \bar B
= B^T~.
\eeq
The couplings at the baryon vertex may be represented by traces $\langle
\ldots \rangle$ of $3 \times 3$ matrices, where we use the notation of
Ref.\ \cite{Rosner:1981np}:
\beq
g_{\bar B_1 E B_2} = \gamma \left[ (1-F) \langle \bar B_1 \{ E, B_2 \} \rangle
+ F \langle \bar B_1 [E,B_2] \rangle - (1-2F) \langle \bar B_1 B_2 \rangle
\langle E \rangle \right]~.
\eeq
The last term enforces nonet symmetry, decoupling $s \bar s$ trajectories
from the nucleon.  One then obtains the nonleptonic S-wave decay amplitudes
(correcting a sign in \cite{Nussinov:1969hp})
\bea
\Lambda_-  & \equiv & A(\Lambda \to \pi^- p) = -(2F+1)x/\sx~,\\
\Sigma_+^+ & \equiv & A(\Sigma^+ \to \pi^+ n) = 0~,\\
\Sigma_-^- & \equiv & A(\Sigma^- \to \pi^- n) = -(2F-1) x~, \\
\Xi_-^-    & \equiv & A(\Xi^- \to \pi^- \Lambda) = (4F-1)x/\sx~.
\eea
Here $x$ is an arbitrary scale factor.  Two relations involving factors
$\mp\s$, between $\Lambda_-$ and $\Lambda_0$ and between $\Xi^-_-$ and
$\Xi^0_0$, follow from the $|\Delta I| = 1/2$ nature of the transition.  
A similar relation between $\Sigma^-_-$ and $\Sigma^+_0$ follows from
isospin and the suppression of $A(\Sigma^+ \to \pi^+ n)$, thus providing a test
of octet dominance or our duality assumption.  The full set of S-wave
amplitudes is summarized in Table \ref{tab:amps}, along with observed values
\cite{Roos:1982sd}.  Also shown are values predicted from a best fit, which
selects $F = 1.652,~x= 0.8605$.

\begin{figure}
\begin{center}
\includegraphics[width=0.95\textwidth]{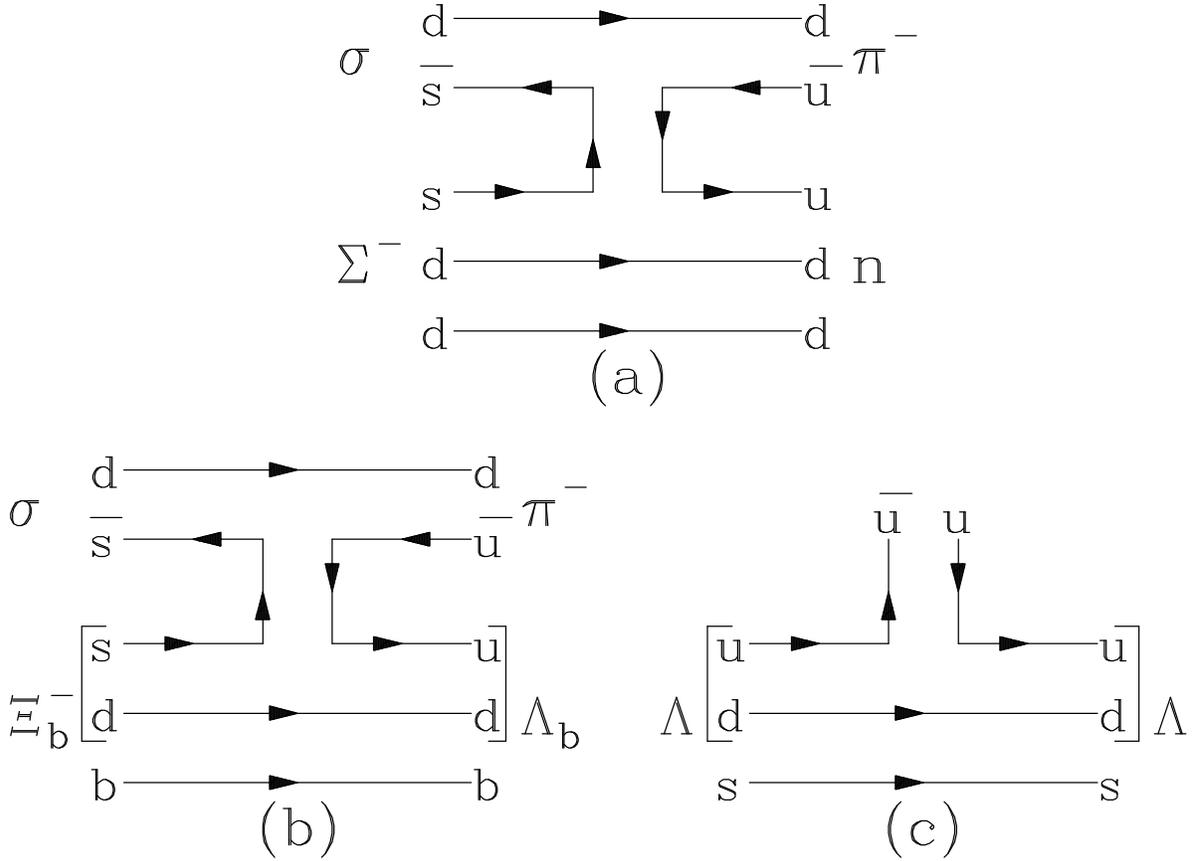}
\end{center}
\caption{Illustration of Reggeon couplings.  (a) Coupling contributing to the
decay $\Sigma^- \to \pi^- n$ induced by a spurion $\sigma$ transforming as a
$K^0$.  (b) Corresponding coupling contributing to the heavy-flavor-conserving
process $\Xi^-_b \to \pi^- \Lambda_b$.  The brackets around the initial $sd$
and final $ud$ pair denote flavor and spin antisymmetry, so the pair are in a
$3^*$ of flavor SU(3) and a state of spin 0.  (c) Corresponding coupling at the
baryon vertex to a quark in hyperons belonging to a flavor- and
spin-antisymmetric state.
\label{fig:coup}}
\end{figure}

% This is Table I
\begin{table}
\begin{center}
\caption{Predicted and observed S-wave amplitudes $A$ for nonleptonic hyperon
decays.  Predicted values are for best-fit parameters $F = 1.652$, $x =
0.8605$
\label{tab:amps}}
\begin{tabular}{c c c c} \hline \hline
     Decay    & Predicted $A$ & Observed & Predicted \\
              &   amplitude   &  value   &   value   \\ \hline
$\Lambda \to \pi^- p$ & $-(2F+1)x/\sx$ & $-1.47 \pm 0.01$ & --1.51 \\
$\Lambda \to \pi^0 n$ & $(2F+1)x/(2\st)$ & $ 1.07 \pm 0.01$ & 1.07 \\
$\Sigma^+ \to \pi^+ n$ &       0      & $0.06 \pm 0.01$ &    0    \\
$\Sigma^+ \to \pi^0 p$ & $-(2F-1)x/\s$ & $-1.48 \pm 0.05$ & --1.40   \\
$\Sigma^- \to \pi^- n$ &  $-(2F-1)x$   & $-1.93 \pm 0.01$ & --1.98   \\
$\Xi^0 \to \pi^0 \Lambda$ & $(4F-1)x/(2\st)$ & $1.55 \pm 0.03$ & 1.39 \\
$\Xi^- \to \pi^- \Lambda$ & $(4F-1)x/\sx$ & $2.04 \pm 0.01$ & 1.97 \\
\hline \hline
\end{tabular}
\end{center}
\end{table}

These relations were also obtained in Refs.\ \cite{Sugawara:1965zza,Suzuki:%
1965zz} within a current-algebra framework.  They imply the triangle relation
\cite{Lee:1964zzc,Sugawara:1964zz}
\beq
2 A(\Xi^- \to \pi^- \Lambda) + A(\Lambda \to \pi^- p) = -
 (3/2)^{1/2} A(\Sigma^- \to \pi^- n)~.
\eeq

\section{RELATION TO $\Xi_b$ DECAY}
\label{sec:rel}

The decay $\Xi^-_b \to \pi^- \Lambda_b$ involves the transformation of a
strange quark in the $\Xi^-_b \simeq b[sd]$ into a $u$ quark in the
$\Lambda_b = b [ud]$ with emission of a pion.  (We use the symbol $\simeq$ to
recall that the light diquark in the $\Xi^-_b$ is nearly, but not totally,
spinless.)  The spectator $b$ quark and the $d$ member of the light diquark
are untouched.  This transition is illustrated in Fig.\ 1(b) as a scattering
process involving the spurion $\sigma$ transforming as a $K^0$.

Note that the upper vertex is the same in Figs.\ 1(a) and 1(b).  We seek a
transition at the light baryon vertex in which the exchanged light-quark
trajectory couples to a spinless $3^*$ of flavor SU(3), as the $u \bar s$
trajectory couples 
at the heavy baryon vertex in Fig.\ 1(b).  Such a coupling is possessed by
the $\Lambda$ when coupling to a $u \bar u$ trajectory, as illustrated in
Fig.\ 1(c).  Here we make use of nonet symmetry, in which the coupling of an
octet strange trajectory to a pair of SU(3) antitriplets is related to the
coupling of an octet-singlet mixture (the $u \bar u$ trajectory) to the
antitriplet pair.  Such symmetry is familiar from the case of the vector
mesons, in which a quark triplet and an antiquark antitriplet form a single
nonet with octet and singlet properties related to one another.

The $3 \times 3$ matrix for the $u \bar u$ exchanged trajectory $E'$ is then
$E' = {\rm Diag}(1,0,0)$.  We calculate $g_{\bar \Lambda E'
\Lambda}$ using the methods of the previous section and find
\beq \label{eqn:xiamp}
A(\Xi^-_b \to \pi^- \Lambda_b) = (5F - 2)x/3 = 1.796 \pm 0.269 \pm 0.539
= 1.796 \pm 0.602~
\eeq
for the best-fit values $F = 1.652,~x = 0.8605$.
The first error of 15\% has been assigned in accord with the estimate
associated with Reggeon dominance \cite{Nussinov:1969hp}.  (The quality of the
fit in Table \ref{tab:amps} is better than that.)  The second error is a
conservative 30\% associated with possible U(3) breaking in comparing exchange
of strange and nonstrange Regge trajectories.  (A more optimistic view of
uncertainties is presented in Sec.\ \ref{sec:opt}.)
Implicit in this relation is the assumption that the properties of the
light diquark are not greatly affected by the nature of the spectator quark
($b$ in Fig.\ 1(b), $s$ in Fig.\ 1(c)).  A certain enhancement of the decay
amplitude of the spatially more compact $\Xi^-_b$ relative to the hyperon
amplitude has been suggested in Ref.~\cite{Li:2014ada} due to an enhanced
short-distance correlation inside the scalar diquark in the heavy baryon.  
The fact that our calculated rate [see Eq.~(\ref{Br}) below] turns out
to be compatible with experiment \cite{Aaij:2015yoy}
provides an {\it a posteriori} validation of our assumptions.

\section{CALCULATION OF DECAY RATE} \label{sec:pred}

We now apply Eq.\ (\ref{eqn:xiamp}) to the rate calculation in Eq.\
(\ref{eqn:rate}), based on the $A$ amplitude alone, using masses from Ref.\                                     
\cite{Agashe:2014kda}: $m_{\Xi^-_b} = 5.7944$ GeV, $m_{\Lambda_b} = 5.6195$
GeV, $m_{\pi^-} = 0.13957$ GeV (from which we derive $q = 0.1038$ GeV),
obtaining
\beq
\Gamma(\Xi^-_b \to \pi^- \Lambda_b) = 8.27 |A|^2 \times 10^{-16}~{\rm GeV}
= (2.67 \pm 1.79) \times 10^{-15}~{\rm GeV}~,
\eeq
where the error is dominated by the assumed 30\% uncertainty in the amplitude
due to U(3) breaking.  With a $\Xi^-_b$ lifetime of $1.56 \pm 0.04$ ps
\cite{Agashe:2014kda}, this corresponds to a branching fraction
\beq\label{Br}
{\cal B}(\Xi^-_b \to \pi^- \Lambda_b) = (6.32 \pm 4.24 \pm 0.16) \times
10^{-3} = (6.3 \pm 4.2) \times 10^{-3}~,
\eeq
where the errors correspond to those of $A$ and the $\Xi^-_b$ lifetime. 
Neglecting the effect of small mass differences, the branching fraction
for $\Xi^0_b \to \pi^0 \Lambda_b$ is then expected to be half of this.

\section{ALTERNATIVE VIEW OF UNCERTAINTIES} \label{sec:opt}

One can express relations for couplings of the strangeness-changing spurion
and $\pi^-$ purely in terms of strangeness-changing observables.  Then one
needs an estimate of the accuracy of U(3) symmetry in relating couplings
of the {\it same} trajectory to different particles.  For SU(3), this was
done long ago by Barger and Olsson \cite{BandO}, with no deviation found to an
accuracy of about 25\%.  However, one can do better if restricting attention
to S-wave nonleptonic hyperon decays.

As $A(\Xi^-_b \to \pi^- \Lambda_b)$ is predicted by our assumption to depend just on the two
parameters $F$ and $x$, one can express it as a linear combination of
any two $A$ amplitudes for nonleptonic hyperon decays.  Specifically, one has
\bea
A(\Xi^-_b \to \pi^- \Lambda_b) &=& -\frac{1}{2\sx}A(\Lambda \to \pi^- p)
- \frac{3}{4}A(\Sigma^- \to \pi^- n) = 1.75~,\\
                               &=& \frac{1}{\sx} [A(\Lambda \to \pi^- p)
+ 3 A(\Xi^- \to \pi^- \Lambda)] = 1.90~,\\
                               &=& -\frac{1}{2}A(\Sigma^- \to \pi^- n)
+ \frac{1}{\sx} A(\Xi^- \to \pi^- \Lambda) = 1.80~,
\eea
inserting the experimental values from Table I. One finds a maximum deviation
of about 6\% from the value using fitted $F,x$.  This 
precision provides a test for SU(3) as the above amplitudes involve different 
members of the baryon octet in the initial and final states. It is well within the 15\%
associated with the expected error from assuming Reggeon dominance.  Adding
amplitude uncertainties of 6\% and 15\% in quadrature and taking account of
the small uncertainty in the $\Xi^-_b$ lifetime, the total uncertainty in
the predicted branching fraction is reduced from $4.2 \times 10^{-3}$ to
$2.0 \times 10^{-3}$, still consistent with the LHCb range but now favoring
its upper end.

\section{CONCLUSIONS} \label{sec:con}

We have shown that an application of Dolen-Horn-Schmid duality to the recently
observed decay $\Xi^-_b \to \pi^- \Lambda_b$ leads to a 
calculated branching fraction ${\cal B} = (6.3 \pm 4.2) \times 10^{-3}$, 
consistent with the LHCb range \cite{Aaij:2015yoy}.
If a more optimistic view of the accuracy of U(3) is adopted, our prediction
would favor the upper end of this range.  It will be interesting to see if this
estimate is borne out when the suggestion of Ref.\ \cite{Voloshin:2015xxa} 
or reducing the uncertainty on $f_{\Xi^-_b}/f_{\Lambda_b}$ is applied.

\section*{ACKNOWLEDGMENTS}

The work of J.L.R. was supported in part by the United States Department of
Energy through Grant No.\ DE-FG02-13ER41598.


\begin{thebibliography}{99}

\bibitem{Cheng:1992ff} 
  H.~Y.~Cheng, C.~Y.~Cheung, G.~L.~Lin, Y.~C.~Lin, T.~M.~Yan and H.~L.~Yu,
  ``Heavy-flavor-conserving nonleptonic weak decays of heavy baryons,''
  Phys.\ Rev.\ D {\bf 46}, 5060 (1992).

\bibitem{Sinha:1999tc} S.~Sinha and M.~P.~Khanna, ``Beauty-conserving
  strangeness-changing two-body hadronic decays of beauty baryons,''
  Mod.\ Phys.\ Lett.\ A {\bf 14}, 651 (1999).

\bibitem{Voloshin:2000et} M.~B.~Voloshin,
``Weak decays $\Xi_Q \to \Lambda_Q \pi$,''
  Phys.\ Lett.\ B {\bf 476}, 297 (2000).

\bibitem{Li:2014ada} X.~Li and M.~B.~Voloshin,
``Decays $\Xi_b \to \Lambda_b \pi$ and diquark correlations in hyperons,''
  Phys.\ Rev.\ D {\bf 90}, 033016 (2014) [arXiv:1407.2556 [hep-ph]].

\bibitem{Faller:2015oma} S.~Faller and T.~Mannel,
``Light-Quark Decays in Heavy Hadrons,''
  Phys.\ Lett.\ B {\bf 750}, 653 (2015) [arXiv:1503.06088 [hep-ph]].

\bibitem{Cheng:2015ckx} 
  H.~Y.~Cheng, C.~Y.~Cheung, G.~L.~Lin, Y.~C.~Lin, T.~M.~Yan and H.~L.~Yu,
  ``Heavy-Flavor-Conserving Hadronic Weak Decays of Heavy Baryons,''
  arXiv:1512.01276 [hep-ph].

\bibitem{Aaij:2015yoy} R.~Aaij {\it et al.} (LHCb Collaboration),
``Evidence for the strangeness-changing weak decay $\Xi_b^-\to\Lambda_b^0
\pi^-$,'' arXiv:1510.03829 [hep-ex].

\bibitem{Maltman:1980er} K.~Maltman and N.~Isgur,
  ``Baryons With Strangeness and Charm in a Quark Model With Chromodynamics,''
  Phys.\ Rev.\ D {\bf 22}, 1701 (1980); H. Lipkin, private communication.

\bibitem{Rosner:1981yh} J.~L.~Rosner,
  ``Magnetic Moments of Composite Baryons, Quarks and Leptons,''
  Prog.\ Theor.\ Phys.\ {\bf 66}, 1422 (1981).

\bibitem{Karliner:2008sv} M.~Karliner, B.~Keren-Zur, H.~J.~Lipkin and
 J.~L.~Rosner, ``The Quark Model and $b$ Baryons,''
  Annals Phys.\ {\bf 324}, 2 (2009) [arXiv:0804.1575 [hep-ph]].

\bibitem{Gronau:1972pj} M.~Gronau,
  ``Nonleptonic hyperon decays in a current-current quark model,''
  Phys.\ Rev.\ D {\bf 5}, 118 (1972) [Phys.\ Rev.\ D {\bf 5}, 1877 (1972)].

\bibitem{Gronau:1972pa} M.~Gronau,
  ``Hadronic hyperon decays in the symmetric quark model,''
  Phys.\ Rev.\ Lett.\ {\bf 28}, 188 (1972).

\bibitem{Wu:1985yb} D.~d.~Wu and J.~L.~Rosner,
  ``Constituent-quark description of nonleptonic hyperon decays,''
  Phys.\ Rev.\ D {\bf 33}, 1367 (1986).

\bibitem{Nussinov:1969hp} S.~Nussinov and J.~L.~Rosner,
  ``Duality and nonleptonic hyperon decay,''
  Phys.\ Rev.\ Lett.\  {\bf 23}, 1264 (1969).

\bibitem{Dolen:1967zz} R.~Dolen, D.~Horn and C.~Schmid,
  ``Prediction of Regge Parameters of $\rho$ Poles from Low-Energy $\pi N$
   Data,''
  Phys.\ Rev.\ Lett.\ {\bf 19}, 402 (1967).

\bibitem{Dolen:1967jr} R.~Dolen, D.~Horn and C.~Schmid,
  ``Finite energy sum rules and their application to $\pi N$ charge exchange,''
  Phys.\ Rev.\ {\bf 166}, 1768 (1968).
  
\bibitem{Sugawara:1965zza} H.~Sugawara,
 ``Application of Current Commutation Rules to Nonleptonic Decay of Hyperons,''
  Phys.\ Rev.\ Lett.\ {\bf 15}, 870 (1965).

\bibitem{Suzuki:1965zz} M.~Suzuki, ``Consequences of Current Commutation
  Relations in the Nonleptonic Hyperon Decays,''
  Phys.\ Rev.\ Lett.\ {\bf 15}, 986 (1965).  

\bibitem{Roos:1982sd} M.~Roos {\it et al.} (Particle Data Group Collaboration),
  ``Review of Particle Properties,'' Phys.\ Lett.\ B {\bf 111}, 1 (1982).  See
  in particular the mini-review by O. Overseth, p.\ 286.

\bibitem{Rosner:1981np} J.~L.~Rosner, ``Graphical Form of Duality,''
  Phys.\ Rev.\ Lett.\ {\bf 22}, 689 (1969).

\bibitem{Lee:1964zzc} B.~W.~Lee,
  ``Transformation Properties of Nonleptonic Weak Interactions,''
  Phys.\ Rev.\ Lett.\  {\bf 12}, 83 (1964)

\bibitem{Sugawara:1964zz} H.~Sugawara, ``A New Triangle Relation for
Nonleptonic Hyperon Decay Amplitudes as a Consequence of the Octet Spurion and
the R Symmetry,'' Prog.\ Theor.\ Phys.\ {\bf 31}, 213 (1964).

\bibitem{Agashe:2014kda} 
  K.~A.~Olive {\it et al.} (Particle Data Group Collaboration),
  ``Review of Particle Physics,'' Chin.\ Phys.\ C {\bf 38}, 090001 (2014),
and 2015 update.

\bibitem{BandO} V. Barger and M. Olsson, Phys.\ Rev.\ Lett.\ {\bf 18}, 294
(1967).

\bibitem{Voloshin:2015xxa} M.~B.~Voloshin, ``Remarks on measurement of the
 decay $\Xi_b^- \to \Lambda_b \pi^-$,'' arXiv:1510.05568 [hep-ph].

\end{thebibliography}
\end{document}